\documentclass[fleqn,usenatbib]{mnras}
\usepackage{newtxtext,newtxmath}
\usepackage[T1]{fontenc}
\usepackage{multirow}
\DeclareRobustCommand{\VAN}[3]{#2}
\let\VANthebibliography\thebibliography
\def\thebibliography{\DeclareRobustCommand{\VAN}[3]{##3}\VANthebibliography}
\usepackage{ragged2e}
\usepackage{multirow}
\usepackage{graphicx}	
\usepackage{amsmath}	
\usepackage{color}
\definecolor{referee}{rgb}{0,0,0}
\usepackage{CJKutf8}

\title[Sudoku at high-redshift]{ High-$z$ Sudoku:
A diagnostic tool for identifying robust (sub)mm redshifts
}

\author[Tom Bakx \& Helmut Dannerbauer]{Tom J. L. C. Bakx$^{1,2}$\thanks{E-mail: bakx@a.phys.nagoya-u.ac.jp (Nagoya University)} and
Helmut Dannerbauer$^{3,4}$
\\
$^{1}$ Division of Particle and Astrophysical Science, Graduate School of Science, Nagoya University, Aichi 464-8602, Japan.\\
$^{2}$ National Astronomical Observatory of Japan, 2-21-1, Osawa, Mitaka, Tokyo 181-8588, Japan. \\
$^{3}$ Instituto Astrof\'isica de Canarias (IAC), E-38205 La Laguna, Tenerife, Spain. \\
$^{4}$ Universidad de la Laguna, Dpto. Astrof\'isica, E-38206 La Laguna, Tenerife, Spain.
}

\date{Accepted XXX. Received YYY; in original form ZZZ}

\pubyear{2021}

\begin{document}
\label{firstpage}
\pagerange{\pageref{firstpage}--\pageref{lastpage}}
\maketitle

\begin{abstract}
We present methods to (i) graphically identify robust redshifts using emission lines in the (sub)mm regime, (ii) evaluate the capabilities of different (sub)mm practices for measuring spectroscopic redshifts, and (iii) optimise future (sub)mm observations towards increasing the fraction of robust redshifts. Using this publicly-available code (\url{https://github.com/tjlcbakx/redshift-search-graphs}), we discuss scenarios where robust redshifts can be identified using both single- and multiple-line detections, as well as scenarios where the redshift remains ambiguous, even after the detection of multiple lines. Using the redshift distribution of (sub)mm samples, we quantify the efficiencies of various practices for measuring spectroscopic redshifts, including interferometers, as well as existing and future instruments specifically designed for redshift searches. Finally, we provide a method to optimise the observation strategy for future (sub)mm spectroscopic redshift searches with the Atacama Large Millimetre/submillimetre Array, where 2~mm proves indispensable for robust redshifts in the $z = 2 - 4$ region.
\end{abstract}




\section{Introduction}
Since their discovery in the 1990's \citep[e.g.,][]{Smail1997,Hughes1998}, the extreme star-formation rates of sub-millimetre galaxies (SMGs) have been difficult to square with our picture of galaxy evolution \citep[see e.g.,][]{blain2002,casey2014}. Their strong emission in sub-mm wavelengths implied star-formation rates in excess of 1000\,M$_{\odot}$/yr, approaching or exceeding the limits of stable galaxy systems through excessive feedback \citep{RowanRobinson2016}, while their strong dust-obscuration poses serious challenges to large-scale unbiased follow-up attempts across all wavelengths \citep{ivison16,Casey2018,Donevski2018,Duivenvoorden2018,bakx2020VIKING,reuter20}. Such large-scale unbiased follow-up is crucial, as studies find SMGs to consist of a diverse population \citep[e.g.,][]{Fardal2001,Baugh2005, Lacey2008,Gonzalez2011, narayanan2015}.

A first step towards such follow-up requires precise and robust redshift measurements. Unfortunately, 
photometric redshifts from sub-mm colours remain only indicative of a redshift range \citep[e.g.,][]{casey2012,pearson13,ivison16,bakx18,Casey2020} due to the degeneracy between dust temperature and redshift \citep{Blain2003}.
Redshifts derived from spectroscopic lines break this degeneracy. 
The first redshift searches used optical spectral lines, however they struggled with the positional uncertainty of SMGs and with a limited bandwidth. Radio observations improved the positions, but provide a strong bias towards lower redshifts, and are not workable for large, complete redshift samples \citep[e.g.,][]{chapman2005}.
Instead, redshift searches using (sub)millimetre spectroscopy are unhindered by dust extinction and can be related unambiguously to the (sub)millimetre source, offering a far better alternative to optical spectroscopy and to the imprecise sub-mm photometric method \citep[e.g.,][]{weiss2009}.
The (sub)mm spectroscopic method has only become competitive with the increased bandwidths of the receivers operating at mm and sub-mm facilities, as for example demonstrated by \cite{Vieira2013} and \cite{weiss2013}, who presented the first redshift survey for 23 strongly-lensed dusty star-forming galaxies selected from the South Pole Telescope (SPT) 2500 deg$^2$ survey using the Atacama Large Millimetre/submillimetre Array (ALMA) at 3~mm. 

\begin{figure*}
	\includegraphics[width=\textwidth]{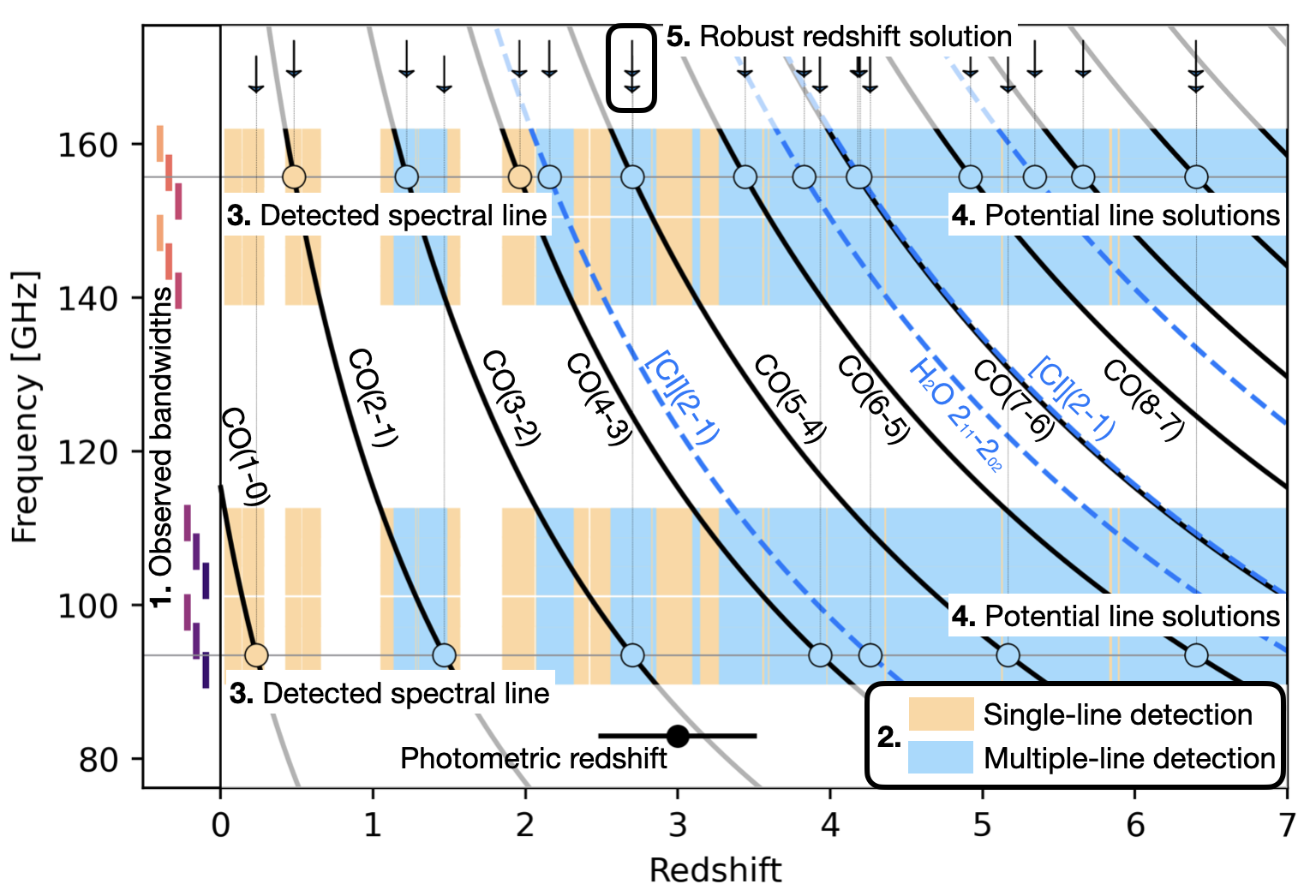}
    \caption{The graphical method reveals the full redshift space ({\it x-axis}) probed by the frequency coverage ({\it y-axis}) of the (sub)mm observations. 
    Assuming the observed bandwidths ({\bf 1.}, {\it shown on the left-hand side}, see Sec.~\ref{sec:OptimizedTunings}), the filled regions show redshifts where single and multiple lines would be detected ({\bf 2.}, {\it orange} and {\it blue fill}, resp.). The horizontal lines ({\bf 3.}) mark the observed lines, indicating the potential redshift solutions where they cross the observable lines ({\bf 4.}, {\it circles}) with the fill matching the underlying colour. The {\it top arrows} ({\bf 5.}) indicate the position of the potential redshift solution of each line, where the arrows line up in the case of a multi-line redshift. The photometric redshift ({\it black errorbar}) can help guide the redshift solution.
    In this example, the two observed CO line transitions, CO(3-2) and CO(5-4), result in a robust redshift identification at $z = 2.7$, since we can exclude all but one other solution where the two arrows do not line up. This other solution at $z \approx 6.4$ can be excluded through the lack of emission at 109.040 and 140.195\,GHz. {\color{referee} Note that the horizontal lines at $\sim 101$~GHz and $\sim 151$~GHz are small (0.5~GHz) gaps in the frequency coverage.}}
    \label{fig:fig1}
\end{figure*}
The large uncertainties of photometric redshifts and the lack of discerning spectral features of (sub)mm spectral lines (i.e., unlike the resonant, a-symmetric emission from Ly$\alpha$) complicate the spectroscopic redshift determination. Instead, the process of robust redshift determination at (sub)mm wavelengths, in particular using rotational emission lines from carbon-monoxide (CO), involves the careful comparison between the potential redshift solutions of individual lines, and then selecting the solution where multiple lines agree. 
This method is far from ideal, as it leads to confirmation bias for solutions close to the $z_{\rm phot}$ \citep{Yanai2020}, and fails to draw attention to redshift degeneracies. Similarly, sometimes redshift solutions can be excluded if we fail to detect additional spectral lines within the bandwidth.
In this paper, we expand on the graphical method from \cite{Bakx2020IRAM} which provides a more user-friendly method for identifying robust redshifts. This method is capable of highlighting redshift degeneracies, and enables the easy removal of untenable redshift solutions, similar to graphical-numerical puzzles, such as the Sudoku\footnote{While invented in the USA, the late Maki Kaji (\begin{CJK*}{UTF8}{min}鍜治  真起\end{CJK*}) popularized the game as {\bf Sū}ji wa {\bf doku}shin ni kagiru (\begin{CJK*}{UTF8}{min}数字は独身に限る\end{CJK*}), roughly translating to {\it the digits must be single}. } puzzle.
We describe the method in detail in Section~\ref{sec:GraphMethod}, where we explain different scenarios and note important caveats. In Section~\ref{sec:redshiftSearches}, we quantify the efficiency of various methods for finding spectroscopic redshifts, and provide a method for finding the optimum tunings for future redshift searches in Subsection~\ref{sec:OptimizedTunings}. The python-based code and example scripts for each of the graphs in this paper are publicly available at \url{https://github.com/tjlcbakx/redshift-search-graphs}, and can be easily and swiftly adapted for the reader's own needs. 

\section{Graphical method}
\label{sec:GraphMethod}
Robust redshifts require the exclusion of any other potential redshift interpretations of the line emission except one.
Numerically solving this cross-comparison is a tedious approach, prone to errors and overlooked options. Instead, we use the following graphical method for identifying the potential spectroscopic redshifts for each line. We further detail several redshift identification examples, and list several caveats with this method specifically and redshift searches in general.

\subsection{Explanation of the graphical method}
In Figure~\ref{fig:fig1}, we show this graphical method for a fiducial $z_{\rm phot} = 3$ example source observed with an efficient ALMA tuning setup (see Section\,\ref{sec:OptimizedTunings}). Here, the observed frequency range ({\it y-axis}) is shown against the redshift range ({\it x-axis}) using the following approach:
\begin{enumerate}
\renewcommand{\labelenumi}{\arabic{enumi}}
    \item {\bf Observed bandwidths:} Indicate the observed bandwidths, in this case shown as both the LSB and USB of the ALMA tunings.
    \item {\bf Single- and multiple-line detection:} The observation is expected to detect the CO-lines (\textit{black lines}), and might detect the [C\,{\sc i}] and H$_2$O lines (\textit{dashed blue lines}; see Section~\ref{sec:excludingRedshiftRegions}). \textit{Orange} and \textit{blue fill} indicate where the observations would be able to observe one or more CO spectral lines, respectively.
    \item {\bf Detected spectral lines:} Each actually-observed spectral line is indicated with a \textit{horizontal line} at the observed frequency. 
    \item {\bf Potential line solutions:} Where the observed lines cross the redshifted spectral lines, a {\it circle} marks the potential spectroscopic redshift, with the circle's fill representing the underlying colour. At the top of the graph, arrows indicate each potential spectroscopic redshift solution.
    \item {\bf Robust redshift solution:} The top arrows line up when multiple lines agree on a robust redshift, identifying the spectroscopic redshift.
\end{enumerate}

This final check is important, because CO-line redshift surveys can suffer from potential multiple redshift solutions (see Section \ref{sec:inconclusiveRedshifts} and Figure~\ref{fig:fig4}), since the CO-ladder increases linearly with each transition J$_{\rm up}$. For example, the observation of J$_{\rm{up}}$ = 2 and 4 CO-line transitions of a z = 2 galaxy (76.7 and 153 GHz, resp.) could also be interpreted as the J$_{\rm{up}}$ = 3 and 6 CO-line transitions of a z $\approx$ 3.5 galaxy. This mis-identification is a non-negligible possiblity given the large 13\% uncertainties in sub-mm derived photometric redshifts of around $\Delta z$\,$\sim$\,1 (defined as $\Delta z / (1+z)$, e.g. \citealt{pearson13,ivison16,bakx18}), although Bendo et al. in prep. finds a sharp reduction in uncertainty down to $\sim 7$\%.

\subsection{Robust redshifts}

\subsubsection{Robust redshifts from CO lines}
Figure~\ref{fig:fig1} shows the derivation of a spectroscopic redshift of $z_{\rm spec} = 2.7$, with two lines at 93.463 and 155.772\,GHz, corresponding to CO(3-2) and CO(5-4) respectively. A high-$z$ redshift solution ($z_{\rm spec} = 6.4$) exists because of the linear spacing between CO line transitions and their isotopologues, but this solution would require additional lines at 109.04 and 140.195\,GHz. The lack of such emission can be used to exclude this redshift possibility.

\subsubsection{Robust redshifts from CO and ancillary lines}
\begin{figure}
	\includegraphics[width=\columnwidth]{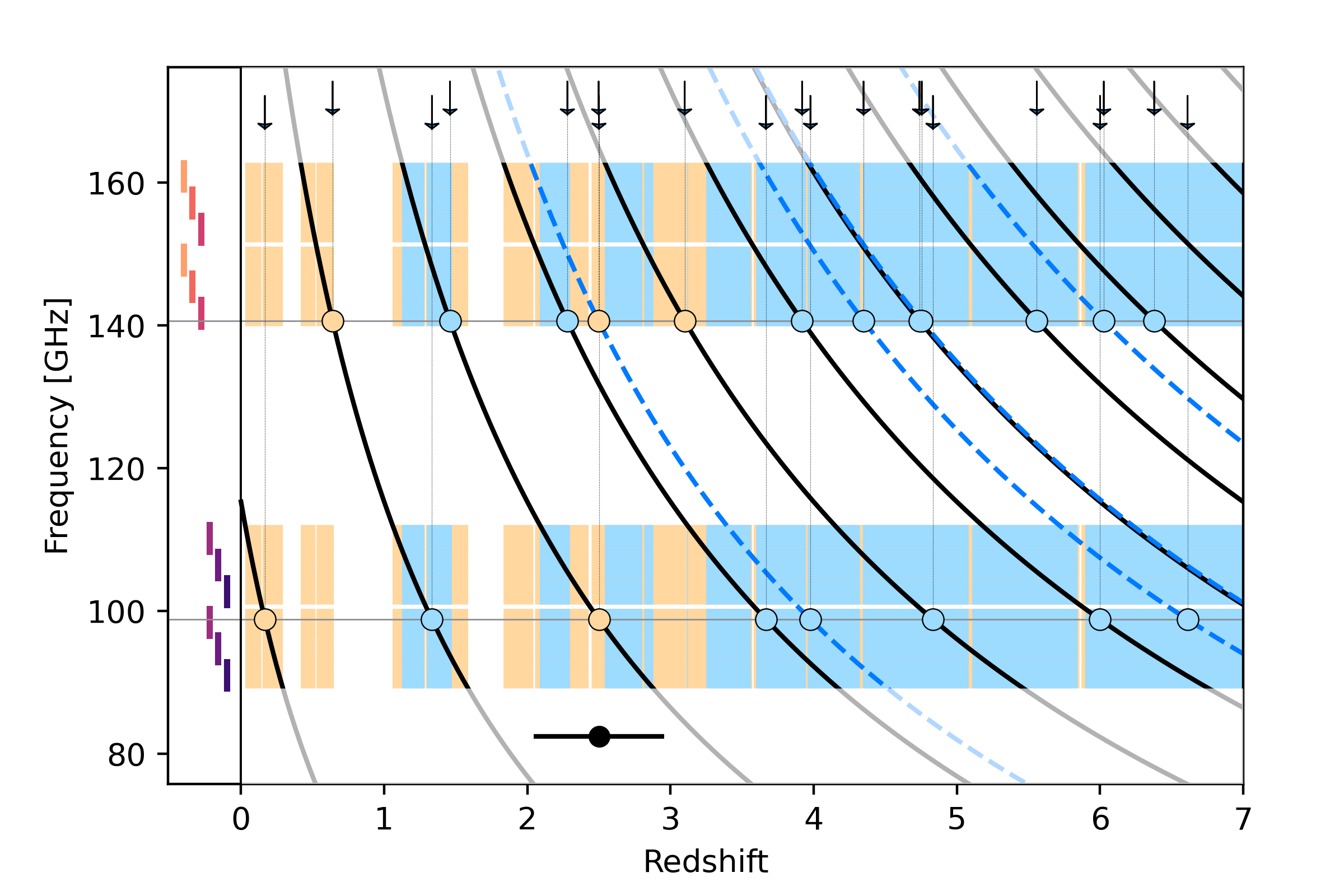}
    \caption{Same as Fig. \ref{fig:fig1}, showing a robust redshift identification at $z_{\rm spec} = 2.5$ using the efficient ALMA configuration through a combination of CO and [C\,{\sc i}] emission (98.804 and 140.607\,GHz resp.). Because non-CO lines do not follow the linear scaling of CO emission lines, any combination of a redshift identification using CO and such additional lines is robust, since it removes the possibility of inconclusive redshifts due to degeneracy in the CO transitions. }
    \label{fig:fig2}
\end{figure}
Non-CO lines do not scale linearly with CO line emission, and thus always offer a robust derivation of the spectroscopic redshift. These lines are typically fainter than the CO emission, however they do offer the ability of near-guaranteed robust redshift identifications (see Sec. \ref{sec:excludingRedshiftRegions}). This is because the high fidelity of the central frequency (typically in excess of $f/\Delta f$ $\approx 10^2 - 10^{4}$) leaves no room for multiple interpretations. In Figure~\ref{fig:fig2}, we show this scenario for a fiducial $z_{\rm spec} = 2.45$ galaxy with observed CO(3-2) and CI(1-0) at 98.804 and 140.607\,GHz respectively. 

\subsubsection{Robust redshifts from single-line detections}
\begin{figure}
	\includegraphics[width=\columnwidth]{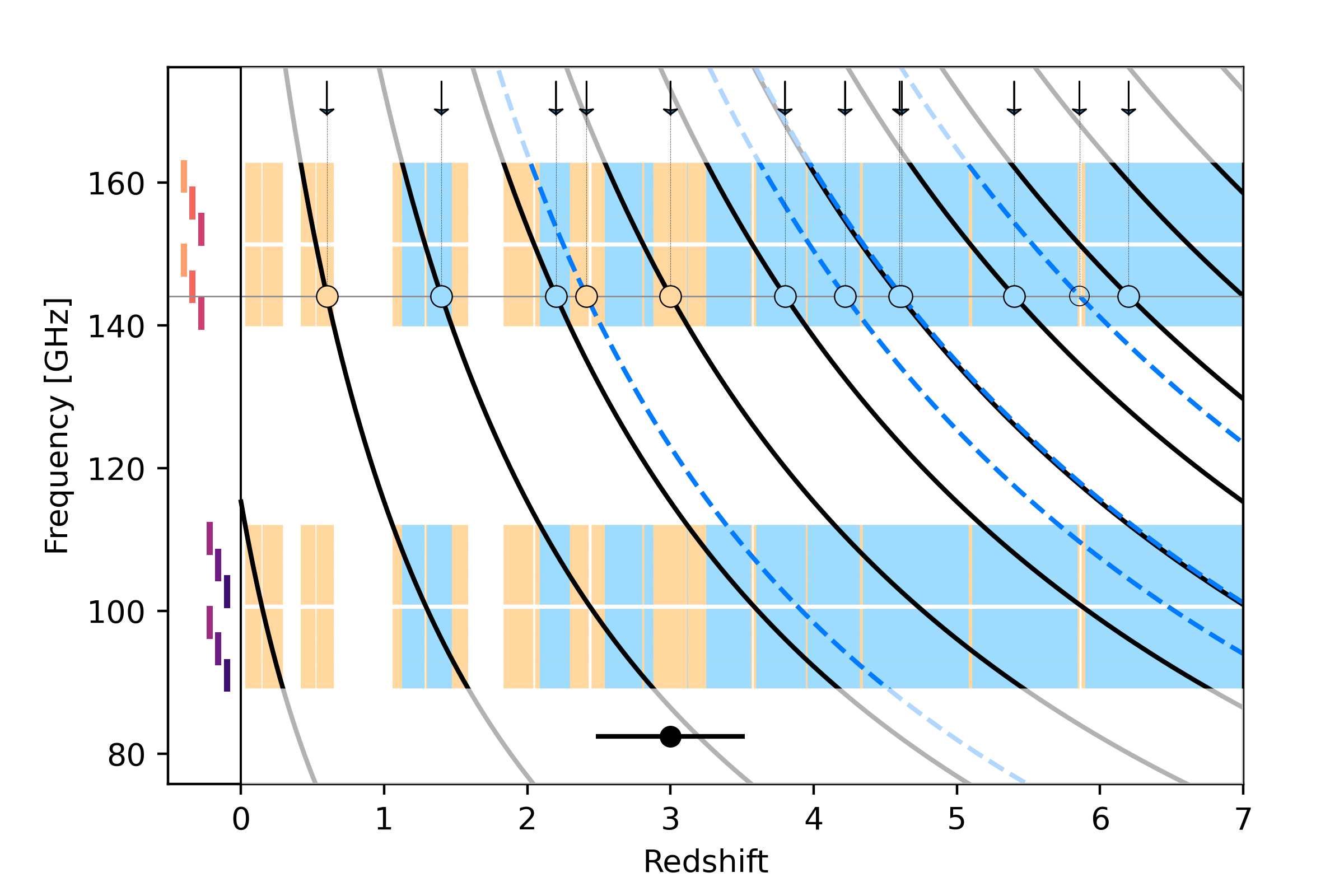}
    \caption{Same as Fig.\,\ref{fig:fig1}, showing a robust redshift at $z = 3.0$ using the efficient ALMA configuration through just a single spectral line. Here, we require that the single spectral line is observed at high significance ($>$8$\sigma$), with no detectable emission at any of the potential line solutions. The low-redshift solution can often times be excluded through optical or near-infrared follow-up (see Section~\ref{sec:excludingRedshiftRegions}), and equation \ref{eq:Nsigma} tells us the $z = 0.6$ solution is 4.5$\sigma$ [$\approx (5-2)/(0.13*5)$] removed using just the sub-mm $z_{\rm phot}$. }
    \label{fig:fig3}
\end{figure}
These graphs allow us to visualize the bandwidth, and associated redshift possibilities that our observations have covered. As shown in Figure\,\ref{fig:fig3}, if we detect a convincingly-bright line (see Section \ref{sec:excludingRedshiftRegions}) at 144.089~GHz using our efficient ALMA tuning (see Section ~\ref{sec:OptimizedTunings}), we can exclude most neighbouring redshift solutions in the same vein as the exclusion of the $z_{\rm spec}= 6.4$ solution in Fig.~\ref{fig:fig1}. We would have detected lines at other frequencies for all but the lowest-redshift solution ($z_{\rm spec} = 0.6$). Using the photometric redshift as a guide, we can exclude this possibility, leaving only $z_{\rm spec} = 3.0$ as a solution, with the detected line being CO(5-4). We note that the $z_{\rm spec} = 2.41$ solution (where the detected line would be [C\,{\sc i}]) would still require a CO line to be detected at 101.251\,GHz.

\subsection{Caveats}
\subsubsection{Inconclusive redshifts from multiple CO lines}
\label{sec:inconclusiveRedshifts}
\begin{figure}
	\includegraphics[width=\columnwidth]{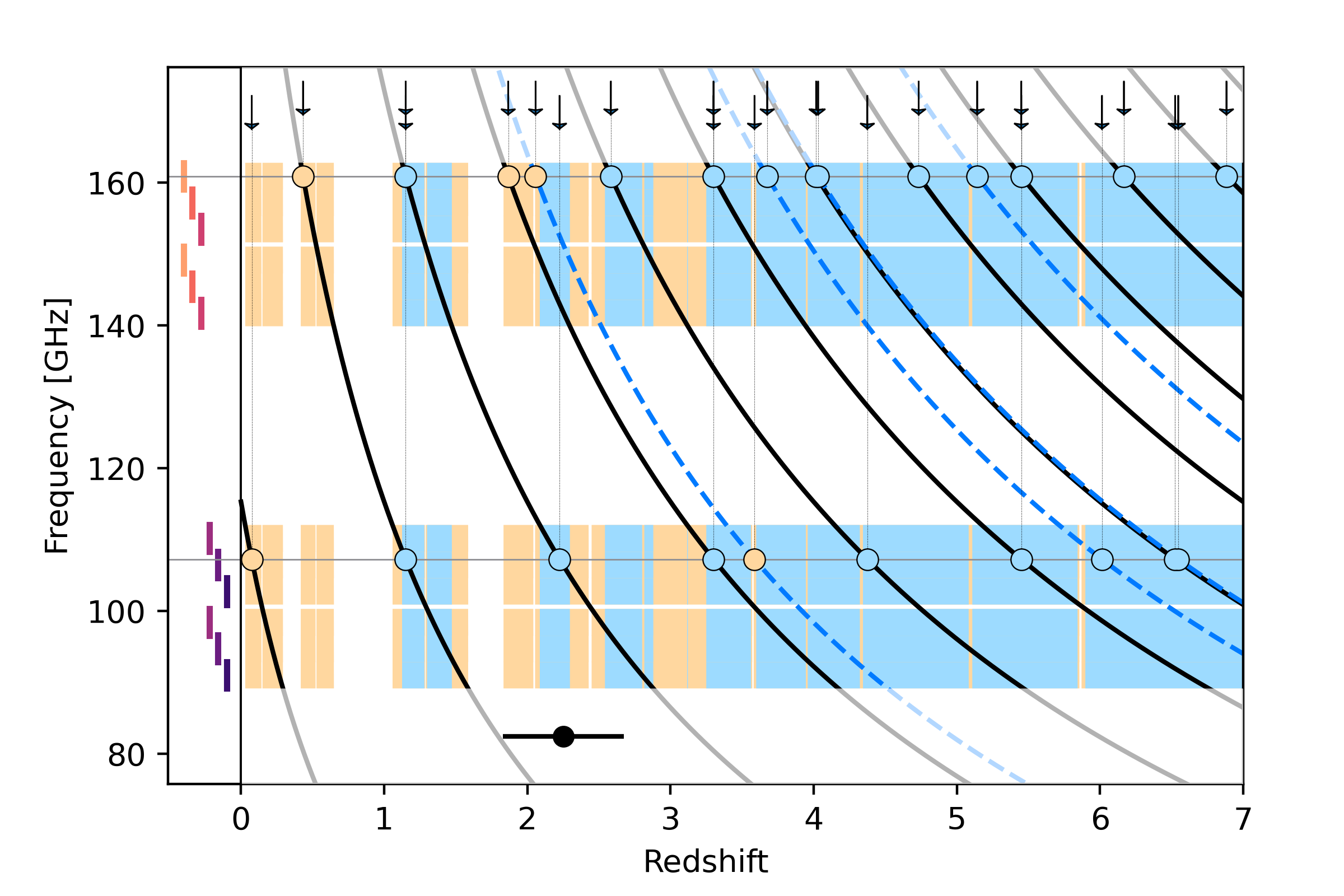}
    \caption{Same as Fig.\,\ref{fig:fig1}, showing an inconclusive redshift from multiple detected CO lines. The detected lines at 107.229 and 160.844\,GHz can correspond equally-well to $z_{\rm spec} = 1.15$ or 3.30. Such scenarios appear when a single scalar multiplied with both CO line transitions results in other integer possibilities. }
    \label{fig:fig4}
\end{figure}
Unfortunately, not all multi-line solutions can be excluded through the non-detection of other lines (such as in Fig.\,\ref{fig:fig1}). We show such a scenario in Figure\,\ref{fig:fig4}, where two redshift solutions of two lines at 107.229 and 160.844\,GHz can correspond to $z_{\rm spec} = 1.15$ or 3.30. The low-redshift scenario would detect CO(2-1) and CO(3-2), while the higher-redshift solution involves CO(4-3) and CO(6-5).
Such scenarios occur when the ratio of the upper transition (J$_{\rm up}$) of both CO line transitions can be multiplied to form another set of integer numbers. In this case, the common multiplier is {\it two}, however extensive possibilities exist. We can quantify the change in redshift between a scenario where we either detected CO($a$, $a-1$) for one redshift solution or CO($b$, $b-1$) for the other redshift solution as follows
\begin{equation}
    \frac{\Delta z}{1+z} = \frac{b-a}{a}. \label{eq:dz}
\end{equation}

The 1$\sigma$ uncertainty in $\Delta z / (1+z)$ is typically on the order of 13\,\% \citep[e.g.,][although this is shown to decrease to only $\sim 7$\% with deep continuum detections at 2~mm by Bendo et al. in prep.]{pearson13}. In case the photometric redshift of a line agrees with a solution of CO($a$, $a-1$), the photometric redshift can thus identify how many standard-deviations ($N$) we are offset from a solution at CO($b$, $b-1$) using
\begin{equation}
    N [\sigma] \approx \frac{b-a}{a \times{} 0.13}. \label{eq:Nsigma}
\end{equation}

In case of multiple detected lines, we can identify the nearest combination of lines that can be confused for one-another. Mathematically, this multiplier can be identified using the greatest common divisor, the $GCD$. The nearest solutions are the multiplication of the remaining components with $GCD \pm 1$. {\color{referee} In the case of a tentative redshift based on CO(4-3) and CO(6-5), the largest common multiplier of the CO transitions (J$_{\rm up} = 4$ and 6, resp.) is 2, leaving additional redshift solutions using CO(2-1) and CO(3-2), as well as CO(6-5) and CO(9-8), typically offset by $\sim 3.8 \sigma$ [$\approx (6 - 3)/(0.13 \times{} 6)$]. Note that equation \ref{eq:dz} and \ref{eq:Nsigma} do not differ on whether one takes the lower- or higher-frequency line, nor the lower- or higher-redshift possibility.}

\subsubsection{Excluding redshift regimes}
\label{sec:excludingRedshiftRegions}
We can exclude redshift regions where we would have seen lines in a deep enough survey. These are typically restricted to CO lines, since the correlation between CO lines and [C\,{\sc i}] emission strongly depends on the particular ISM conditions (see e.g., \citealt{Valentino2018}). CO lines transitions, instead, are strongly correlated to one-another through transitional partition functions following so-called spectral line energy distributions (SLEDs). The Milky Way CO SLED has a sharp drop-off in velocity-integrated fluxes for J$_{\rm up}$ transitions greater than 3, however for star-bursting galaxies, these SLEDs can approach the {\it maximum} thermalized profile, where velocity-integrated fluxes increase with $J^2$ (see \citealt{carilli2013}). 

\begin{figure}
    \centering
    \includegraphics[width=\linewidth]{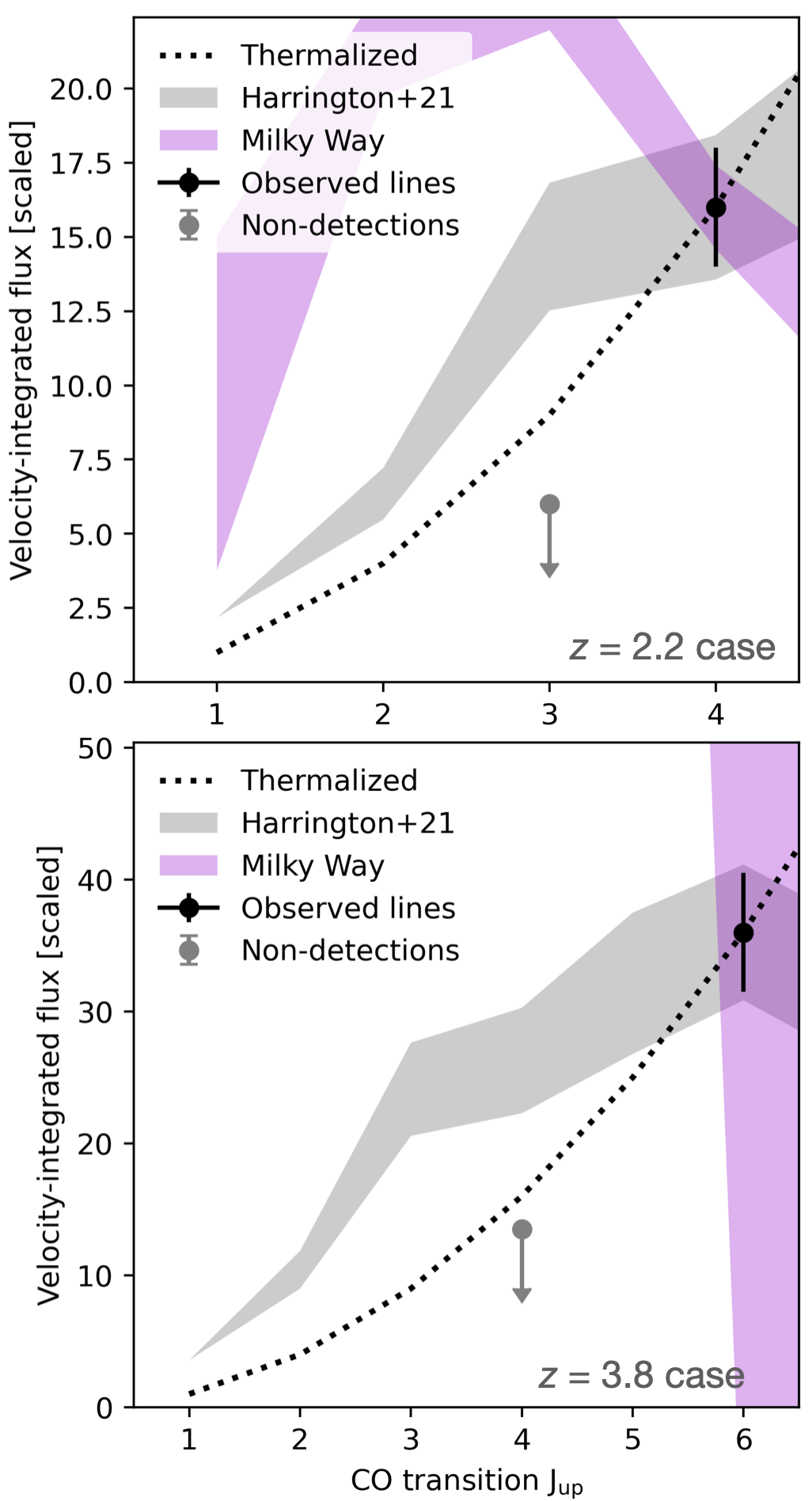}
    \caption{{\color{referee}We compare the CO SLEDs of the alternative multi-line redshift solutions shown in Figure~\ref{fig:fig3}, with the lower-redshift $z = 2.2$ case in the {\it top panel} and the higher-redshift $z = 3.8$ case in the {\it bottom panel}. These graphs inform us whether we can exclude redshift possibilities based on the upper-limits from deep observations. In the lower-redshift scenario, we would have detected CO(4-3) at 8$\sigma$ and not detected CO(3-2) with a similar observation depth. In the higher-redshift scenario, we would have detected CO(6-5) at 8$\sigma$ and not detected CO(4-3). We compare against various SLEDs. The CO SLEDs from \citet{Harrington2021} in {\it grey fill} represents the average SLED from bright {\it Planck}-selected galaxies. The Milky Way inner disk CO SLED from \citet{Fixsen1996} represents a typical low-excitation CO SLED. Finally, the thermalized CO SLED represents the typical maximum expected excitation. In both scenarios, we find the 3$\sigma$ upper limit ({\it grey marker}) below the thermalized profile. This indicates that we can exclude these redshift scenarios, as even a thermalized profile would not be able to produce the flux observed at the higher-J side of the SLED. }
    }
    \label{fig:Fig5}
\end{figure}

In the case of Fig.\,\ref{fig:fig1}, we know the undetected CO lines should be at least on the same order of brightness as the detected lines, as their transitions are in between the detected emission lines. For the single-line case in Fig.\,\ref{fig:fig3}, the $z_{\rm spec} = 2.2$ case would have, in the most unfortunate case, a CO(3-2) line with an integrated flux of $\sim 56$\% (= 3$^2$/4$^2$) fainter than the detected line. Similarly, for the higher-redshift solution at $z_{\rm spec} = 3.8$, the CO(5-4) would be $\sim 70$\% (= 5$^2$/6$^2$) as bright as the detected line. {\color{referee} We show this analysis in graphical form in Figure~\ref{fig:Fig5}. }

{\color{referee} Taking into account measurement errors, as well as this line ratio, the exclusion method thus requires relatively-high signal-to-noise ($>8 \sigma$) detections to be robust. In Figure~\ref{fig:Fig5} we show that such a 8$\sigma$ detection of a single spectral line is enough to exclude the potential redshift solution. We show the CO SLED of typical bright DSFGs from {\it Planck}-selected sources \citep{Harrington2021}, a CO SLED from the Milky Way \citep{Fixsen1996}, and the maximum expected excitation from a thermalized profile. We choose the {\it Planck}-selected CO SLED because it traces a large range of CO-transtions, from CO(1-0) to CO(12-11), while still being broadly representative of the population of DFSGs. The Milky Way CO SLED represents typical low-excitation gas, with a steep drop-off, while the thermalized profile reflects the maximum expected excitation. For both the high- and low-redshift possibilities, the upper-limit lies below the thermalized profile. As such, no typical SLED would be able to produce both a non-detection at a low-J, and a detection at a higher-J CO line.
}

\subsubsection{Disqualifying low-redshift interlopers}
As a general rule, redshift deserts become more prevalent at lower redshifts (see the discussion in Section~\ref{sec:redshiftSearches}). Observed frequencies of lines lie far apart, and relatively-narrow bandwidths cover only small $\Delta z$. For example, at $z = 0.5$, CO(1-0) is currently not accessible with ALMA and CO(3-2) requires band 6 ($\sim 345$~GHz) observations, while at $z = 4$ both CO(4-3) and CO(5-4) are covered in band 3 (3~mm; $\sim 100$GHz). 
As such, low-redshift interlopers can only be excluded directly using large observed bandwidths. Another solution involves multi-wavelength observations, which become capable of detecting even heavily dust-obscured galaxies themselves and provide much more accurate photometric redshifts through optical and near-infrared spectral template fitting.

\section{Redshift searches}
\label{sec:redshiftSearches}
\begin{table}
    \caption{Instruments used for redshift searches}
    \label{tab:efficiencies}
    \centering
    \begin{tabular}{rccc}\hline \hline
    Instrument & \multicolumn{3}{c}{Observing bandwidth}  \\
    & [GHz] & / & [mm]  \\ \hline
    ALMA: & $89.1 - 112.0$ \& &\multirow{2}{*}{\LARGE /}& $3.37 - 2.68$ \& \\
     B3 \& B4 & $139.9 - 162.7$ & & $2.14 - 1.84$ \\
    Spectral scan: & $84.2 - 114.9$ \& &\multirow{2}{*}{\LARGE /}& $3.56-2.61$ \& \\
     B3 \& B4   & $125 - 159.25$ & & $2.40 - 1.88$\\
    ALMA B3 fill & $84.2 - 114.9$ &/& $3.56-2.61$ \\
    RSR & $73-111$ &/& $4.11 - 2.70$ \\
    Zpectrometer & $25.6-37.7$ &/& $11.7-7.9$ \\
    Z-Spec & $185-305$ &/& $1.62-0.98$ \\
    DESHIMA & $220-440$ &/& $1.36-0.68$ \\
    Superspec & $200 - 300$ &/& $1.50 - 1.00$\\
    OST (Band 6) & $509 - 893$  &/& $0.59 - 0.34$ \\ \hline \hline
    \end{tabular}
\raggedright \justify \vspace{-0.2cm}
\textbf{Notes:} The top two rows together show the optimised tuning from Figure \ref{fig:band34_optimization}. The third and fourth rows together show the spectral scan set-up, assuming five stacked tunings at both 3 and 2~mm. 
{\color{referee} Note that there exist a small gap in frequency coverage ($\sim 0.5$~GHz) around $\sim 101$~GHz and $\sim 151$~GHz in the optimum ALMA Band 3 \& 4 configuration.}
\end{table}
The graphical method facilitates the analysis once the observations are done. This section complements this by providing a figure of merit for past, existing and future redshift searches (Sections~\ref{sec:interferometricredshiftsearches}, \ref{sec:dedicatedZreceivers}, and \ref{sec:FutureInstrumentation}, respectively). This method can then be used to optimise future redshift searches (see Section~\ref{sec:OptimizedTunings}), and guide the design of instruments for future redshift surveys.

\begin{figure*}
    \centering
    \includegraphics[width=\textwidth]{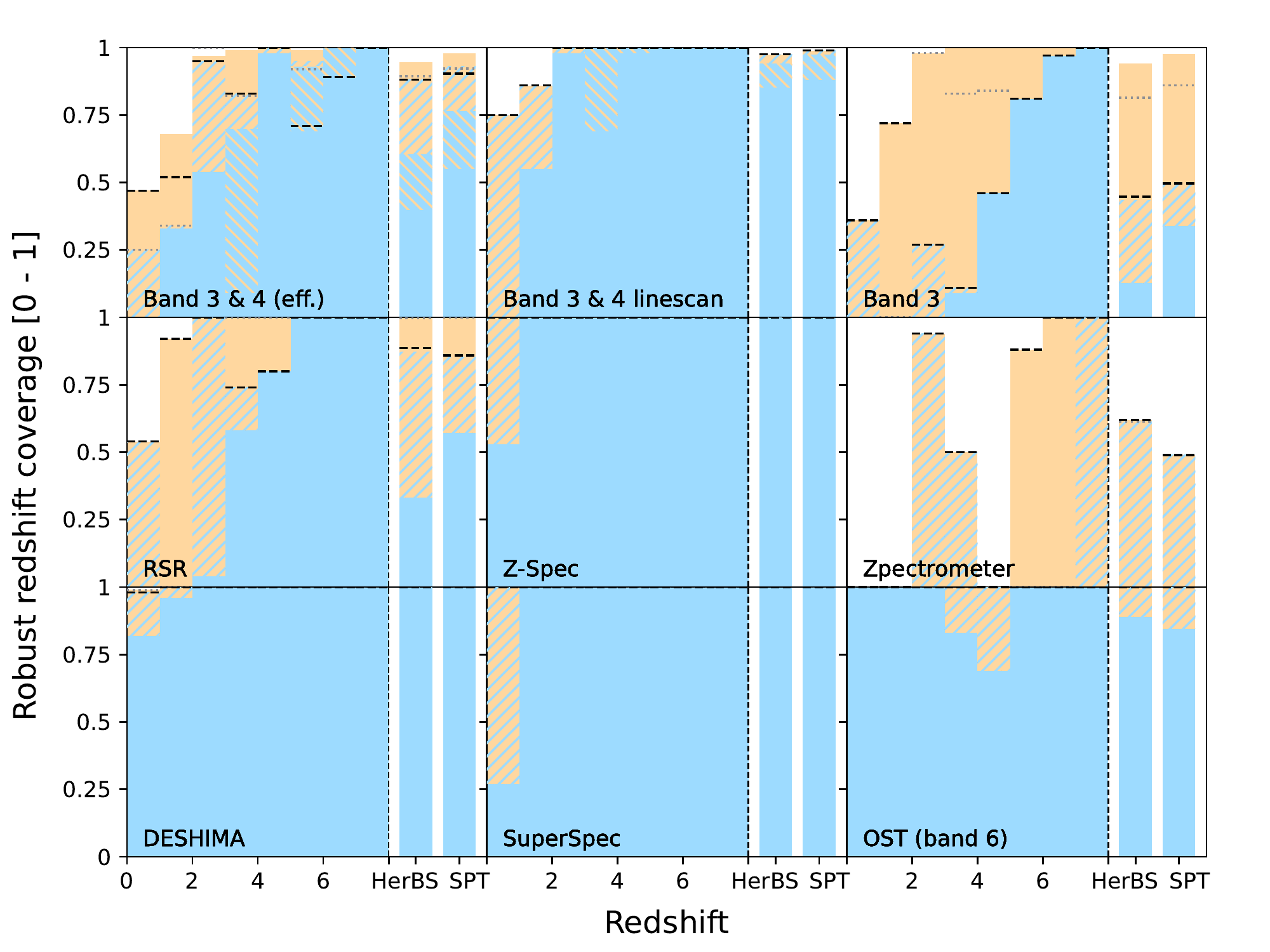}
    \caption{The redshift identification probability as a function of redshift between $z = 0$ and 8, and for smoothed redshift distributions based on two samples (i.e., HerBS; \citealt{bakx18,Bakx2020Erratum}, SPT; \citealt{reuter20}). {\it Orange bars} indicate the fraction of sources for which one would detect a single spectral line, while {\it blue bars} indicate the fraction where two or more spectral lines are detected. {\it Hatched blue fill} indicates the cases where one can identify the redshift robustly with even a single spectral line, while {\it hatched orange fill} indicates the situation where redshift degeneracies remain down to a 5$\sigma$ uncertainty in $z_{\rm phot}$. The {\it grey dotted lines} indicate the robust (single + multiple lines) redshifts when one expects to detect [C\,{\sc i}], and the {\it black dashed lines} indicate the robust (single + multiple lines) if we only assume the improved $z_{\rm phot}$ uncertainties shown in Bendo et al. in prep.
    The {\it top panels} indicate the redshift capabilities of various ALMA set-ups, showing an optimized Band 3 \& 4 set-up (see Table \ref{tab:efficiencies} and Figure \ref{fig:band34_optimization}), a linescan Band 3 \& 4 set-up, and the filled Band 3 tuning from \citet{weiss2013}. The {\it middle panels} show known redshift receivers (RSR; \citealt{Erickson2007}, Z-Spec; \citealt{Naylor2003}, Zpectrometer; \citealt{Harris2007}), and the {\it bottom panels} show the redshift capabilities of future instruments that are currently under development (DESHIMA 2.0; \citealt{Taniguchi2021,Rybak2021}, SuperSpec; \citealt{Redford2021}, Origin Space Telescope (band 6); \citealt{Bradford2021}). }
    \label{fig:redshiftSearchEfficiency}
\end{figure*}

\begin{figure}
    \centering
    \includegraphics[width=\linewidth]{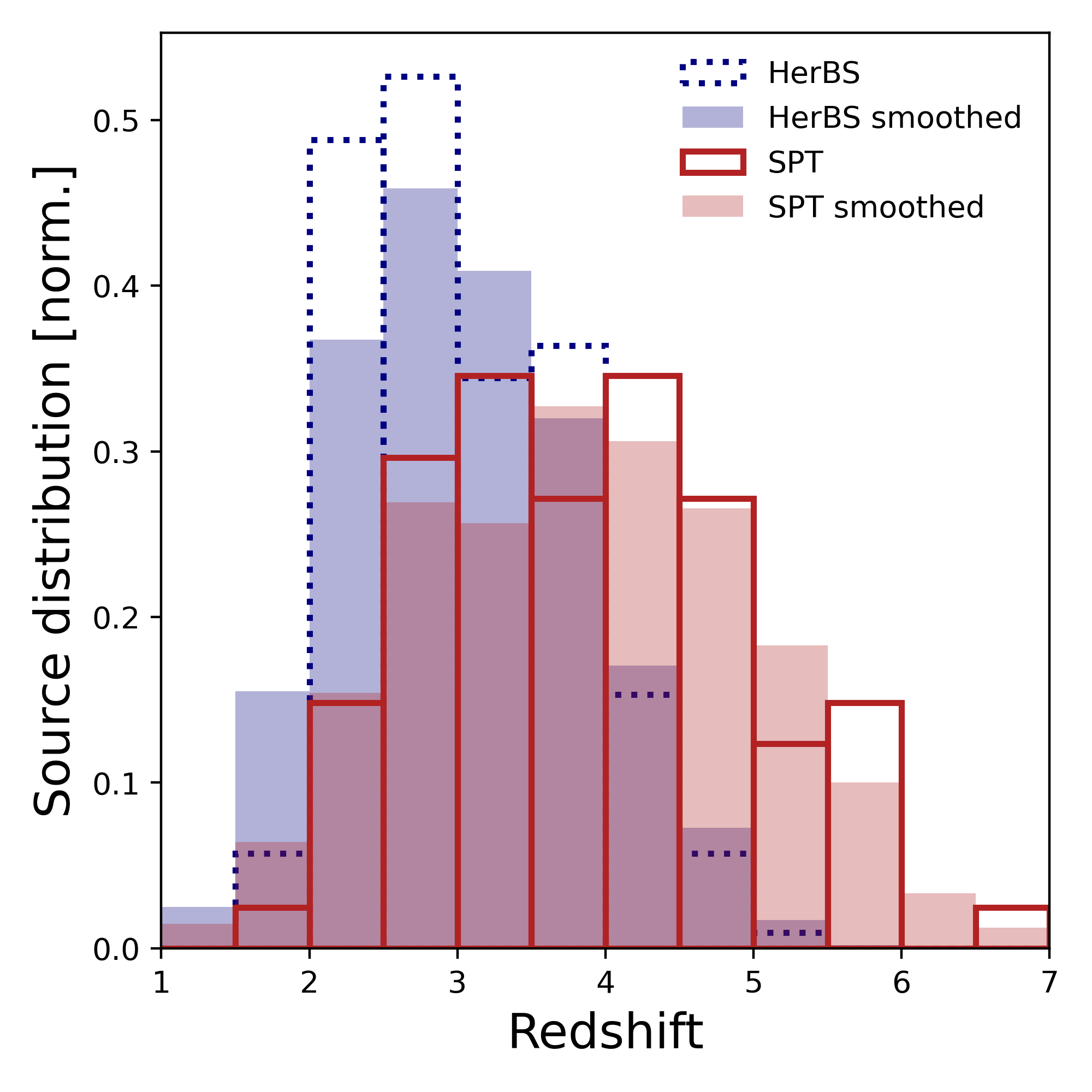}
    \caption{{\color{referee}The redshift distribution of HerBS (\citealt{bakx18,Bakx2020Erratum}) and the South Pole Telescope (\citealt{reuter20}) are used to estimate the efficiency of various observing strategies, shown in {\it dashed lines}. We smooth the reported redshift distribution by around $dz \approx 0.5$ (shown in {\it filled histograms}) in order to avoid any small-number variations in our final efficiency estimations. Each distribution is normalized to a total area of 1. }}
    \label{fig:zdist}
\end{figure}

Figure \ref{fig:redshiftSearchEfficiency} shows the probability of a robust redshift of various configurations and instruments that are often used for redshift searches between $z = 0$ and 8, and for two different DSFG samples (HerBS; \citealt{bakx18,Bakx2020Erratum}, South Pole Telescope; \citealt{reuter20} {\color{referee}, see Fig.\,\ref{fig:zdist}). The average redshift of the {\it Herschel}-selected HerBS sample is $\bar{z} \approx 2.5$, while the SPT sample is selected at longer wavelengths and has an average $\bar{z} = 3.5$. Redshift searches on other DSFG samples, such as ALESS \citep{cunha2015ALESS} or the Ultrared sources \citep{ivison16}, can then be scaled accordingly. The distributions are smoothed by $dz \approx 0.5$ to prevent small-number variations in the final efficiency metrics.} The details of each instrument are shown in Table\,\ref{tab:efficiencies}. For the calculation in Figure~\ref{fig:redshiftSearchEfficiency}, we remove regions that are opaque under normal observing conditions (assuming precipitable water vapour $> 0.5$~mm, with transmission greater than 80\% of the peak). 
We show the total number of sources detected robustly with multiple lines and with single lines in {\it blue fill and hatches}, respectively. The {\it orange fill and hatches} show where the redshift remained ambiguous with one or two lines, while the white region provides a measure of the redshift desert. {\color{referee} 
We define a robust detection by either a detection in CO where all adjacent solutions within 5$\sigma$ (eq.~\ref{eq:Nsigma}) can be excluded, or through the detection of a non-CO line.
}

\subsection{Interferometric redshift searches}
\label{sec:interferometricredshiftsearches}
Interferometers are important tools for finding spectroscopic redshifts \citep[e.g.,][]{Vieira2013,weiss2013,neri2019,Chen2022,Urquhart2022}. By combining multiple spectral windows, interferometers can cover large redshift ranges while delivering high-resolution imaging of both line and continuum emission. However, the telescope array with the largest collecting area, ALMA, only offers a modest spectral coverage (up to $\sim$7.75~GHz per tuning). While the necessary integration time to detect a line from, for example, a bright {\it Herschel} source is only several minutes, an ALMA-based redshift search could thus result in large overheads ($\sim 30$~min.) due to the need to switch tunings. Generally, there are two ways around the large overheads. Firstly, ALMA offers the ability to observe sources close on the sky (typically $\sim10$ degrees of separation) without the need to recalibrate. Observations can then be carried out in batches of sources, reducing the overheads per source \citep[see e.g.,][]{Urquhart2022}. Secondly, ALMA recently started to offer\footnote{https://almascience.eso.org/proposing/proposers-guide} spectral scans that calibrate all spectral windows on the calibrator, cutting down significantly on slewing overheads, which can amount to half or more of the total observing time. If the targeted number of sources within a 10~deg. field is modest (typically $\lesssim 3$), the spectral scan is preferred. Above this, the observation in batches outperforms the spectral scan.

\subsubsection{Filling the 3~mm band}
The initial observations of SPT sources using the full 3~mm band of ALMA successfully demonstrated the interferometer's redshift search capabilities \citep{Vieira2013,weiss2013,strandet2016,reuter20}. These observations use five tunings with ALMA, starting at 84.2~GHz up to 114.9~GHz with overlapping coverage of around one tuning between $96.2 - 102.8$~GHz. 
Figure~\ref{fig:redshiftSearchEfficiency} shows that, while these observations are able to guarantee a single-line detection above $z>2$, this tuning provides only moderate ability to derive robust, unambiguous redshifts on the CO emission alone. For this, 2~mm observations are needed (see e.g., \citealt{Bakx2020IRAM}). [C\,{\sc i}] and H$_2$O emission, as well as improved photometric redshifts can vastly improve the ability of this configuration to estimate the redshift, particularly for samples biased towards high redshifts. Improved uncertainties in the photometric redshift ({\it black dashed lines}) do not improve the redshift capabilities significantly.
Meanwhile, these observations target the CO emission up to the CO(6-5) rotational transition (with one exception up to CO(7-6); \citealt{Marrone2018}). It is good to note that these emission lines correlate with the far-infrared luminosity -- which is often a direct selection function for high-redshift DSFGs --  and these observations are thus less likely to bias observations. Instead, observations aiming at higher-J transitions might be particularly-sensitive to the warmer and denser molecular gas typically found in AGN (e.g., \citealt{Xu2014}).

\subsubsection{Combined 3 and 2~mm tunings}
In Section~\ref{sec:OptimizedTunings} and Figure~\ref{fig:band34_optimization} we describe how to find the optimum tuning for spectroscopic redshifts using ALMA. Figure~\ref{fig:redshiftSearchEfficiency} shows the effect of including band 4 (2~mm). While there exists some regions where multiple lines fail to provide robust redshifts assuming the {\it Herschel}-based photometry \citep[e.g.,][]{pearson13}, the refined photometric redshifts from the 2~mm continuum ({\it black dashed lines}) result in high ($> 70$\%) robust redshifts beyond $z > 2$, at the limited cost of including an additional tuning ($\sim 20$\% longer observation). 
The benefit is particularly stark at the low-redshift end of the galaxy distribution, without needing to rely on the unpredictable [C\,{\sc i}] or H$_2$O lines in redshift searches. These results are in line with the impressive redshift search efficiency shown in \cite{Urquhart2022} (as well as in subsequent follow-up; P.I.: Bakx, 2019.2.00155.S), showing a robust redshift ratio of 72\%, and even promising up to 93\% with deeper observations.

\subsubsection{ALMA spectral scans}
Recent development at ALMA allowed for efficient observations of multiple (up to five) spectral windows for a single source. By calibrating all spectral windows in one go, this observing strategy cuts down significantly on the overheads. We note here that a scan of just band 3 results in similar redshift coverage issues as seen in Figure~\ref{fig:redshiftSearchEfficiency}, as well as the large overlapping coverage seen in e.g., \cite{weiss2013}. The coverage issue, however, can be mitigated by including a spectral scan in band 4, which vastly improves the redshift coverage across a wide range of redshifts. If this is executed as a spectral scan, the overheads will be only relatively minor, however this method will be difficult to execute using batched observations similar to \cite{Urquhart2022}, requiring a doubling in observation time relative to the 3~mm band method.

\subsubsection{NOEMA redshift searches}
Unlike ALMA, NOrthern Extended Millimeter Array (NOEMA; e.g., \citealt{neri2019}) and the IRAM 30m (e.g., \citealt{Bakx2020IRAM}) offer the ability to adjust the observing frequencies dynamically, based on observed spectral lines in other tunings. This is a significant advantage when looking for redshifts, further aided by the wider bandwidths (up to $\sim 16$\,GHz in one tuning, with possible expansion in the future) that allow for a wide 3~mm scan in just two tunings. 
In the 3~mm window, this would amount to an observation similar to the 3~mm scan in Figure~\ref{fig:redshiftSearchEfficiency}, where one typically requires additional observations in the 2~mm band. Follow-up observations in the 2~mm band do not lend themselves for easy generalization, which is why we were not able to generate a similar plot for a NOEMA 3- and 2~mm redshift search, although simply covering the 3- and 2~mm windows similar to the Band 3 \& 4 linescan suggest a near-complete redshift coverage of the HerBS and SPT samples alike using four tunings. Nevertheless, the graphical method of this paper (i.e., Fig.~\ref{fig:fig1}) can offer significant aid in determining additional observations at both 2~mm and other bands, while also highlighting redshift degeneracies.

\subsection{Dedicated redshift receivers}
\label{sec:dedicatedZreceivers}
The advent of spectrometers with instantaneous wide-band enabled the first large-scale redshift surveys. While the first (sub)mm redshift searches involved heterodyne instruments that require multiple tunings (e.g., \citealt{weiss2009}), the notable instruments that performed these dedicated redshift searches are the Redshift Search Receiver (RSR;  \citealt{Erickson2007}), Z-Spec \citep{Naylor2003}, and Zpectrometer \citep{Harris2007} on single-dish telescopes. 

\subsubsection{Redshift Search Receiver}
The Large Millimeter Telescope (LMT)-based observations with the RSR \citep{Erickson2007} combine the instantaneous coverage of the 3~mm window with the powerful 32 to 50~m diameter of the telescope to reveal the spectral lines of galaxies out to high redshifts (e.g., \citealt{Zavala2015,Zavala2018}). At lower redshifts particularly, these observations require ancillary data -- such as with IRAM or $8-10$~m telescopes -- to confirm the redshifts robustly. The addition of a 2~mm spectroscopic ability at the LMT would significantly increase the telescope's ability to determine redshifts robustly (e.g., Kawabe et al. in prep).

\subsubsection{Z-Spec}
The wide bandwidth capability of Z-Spec \citep{Naylor2003} at 1.0 to 1.6~mm enabled the near-guaranteed ability to detect more than one spectral line. As seen in \cite{lupu2012}, the higher-frequency transitions (often of CO lines) are not guaranteed. In this case, clever stacking of the line emission as a function of redshift enabled robust redshifts even though individual lines might be detected only at a few sigma. It is good to note that, unlike the RSR and Zpectrometer, the diameter of the Caltech Submillimeter Observatory (CSO) where the telescope operated was modest at 10~m, since high surface accuracy was needed given the higher-frequency observations.

\subsubsection{Zpectrometer}
The Zpectrometer receiver \citep{Harris2007} operated on the 100~m Green Bank Telescope, and principally targeted CO(1-0) and CO(2-1). This large collecting area is crucial, since the expected emission from these low-excitation rotational transitions is faint \citep{harris2010}. Observationally, the GBT/Zpectrometer set-up had another advantage to increase its efficiency. While performing on-off chopping to remove background and atmospheric noise, the telescope chopped between two galaxy positions. The emission line from the second ('off') source would then be seen as a negative signal \citep{Frayer2011,harris2012}. This method allows for the easy identification of redshifts between 2 and 3.5, as well as above 5. Interestingly, the typical uncertainty on photometric redshifts increases with redshift, going as $\sim 0.13 \times{} (1+z)$, the lower redshifts are basically robust (hatched blue), while the higher redshifts might be mis-interpreted CO(2-1) line emission that is instead CO(1-0).

\subsection{Future instrumentation}
\label{sec:FutureInstrumentation}
New instrumentation is underway to explore the high-redshift Universe using broad-band spectrometers, in particular using Microwave Kinetic Inductance Detectors (MKIDs; \citealt{Day2003,Endo2019}) technology, and on-board of the Origin Space Telescope (OST; \citealt{Battersby2018}). 

\subsubsection{DESHIMA}
The DEep Spectroscopic HIgh-redshift MApper instrument combines an integrated super-conducting spectrometer with a read-out based on MKIDs to enable ultra-wideband ($>100$~GHz) spectroscopy at medium resolution ($f/ \Delta f \approx 500$). We carefully exclude frequency regions with strong atmospheric absorption from our analysis in Figure~\ref{fig:redshiftSearchEfficiency}. As shown in detail in \cite{Rybak2021}, the wide bandwidth at 1.4 to 0.7~mm means that DESHIMA will be sensitive to mid- to higher-J transitions for sources at $z < 3.3$. These emission lines vary strongly from source to source, however can be brought out through similar stacking as seen in Z-Spec (e.g., \citealt{lupu2012}). Beyond $z > 3.3$, the [C\,{\sc ii}] emission line will dominate the emission and will be readily detected. We note that the robust redshift coverage of DESHIMA at even the lowest redshifts is high, and DESHIMA could thus additionally be well-suited towards low-$z$ spectroscopic redshift searches.

\subsubsection{SuperSpec}
The SuperSpec instrument aims to use a lithographically-patterned filterbank (similar to DESHIMA) to probe the high-redshift Universe, primarily through atomic lines of high-redshift targets between $z = 4 - 8$ \citep{Redford2021}. The instrument is designed for operation on the LMT, whose large collecting area allows for deep integration on even faint targets. Similar to DESHIMA, this instrument also has the ability to rapidly assess the redshifts of lower-$z$ targets through mid-J emission lines.

\subsubsection{Origin Space Telescope}
We chose to only model the longest-wavelength band of the medium-resolution spectroscopic instrument on the  OST (band 6; $0.34 - 0.59$~mm, \citealt{Battersby2018}) since it is more sensitive to lower-frequency lines such as CO. We note that additional bands (which are expected to be able to observe simultaneously) improve the redshift capabilities further by picking up ancillary atomic lines which further aid in the determination of robust redshifts. Beyond $z > 1.1$, OST will be able to target [C\,{\sc ii}] emission, leaving only a little gap ($2.73 < z < 2.81$) where at band 6 is unable to pick up either [O\,{\sc iii}]88\micron{} or [C\,{\sc ii}]158\micron{}, although [N\,{\sc ii}]122\micron{} could also be observed.

\subsection{Optimising ALMA tunings for spectroscopic redshifts}
\label{sec:OptimizedTunings}
\begin{figure}
    \centering
    \includegraphics[width=\linewidth]{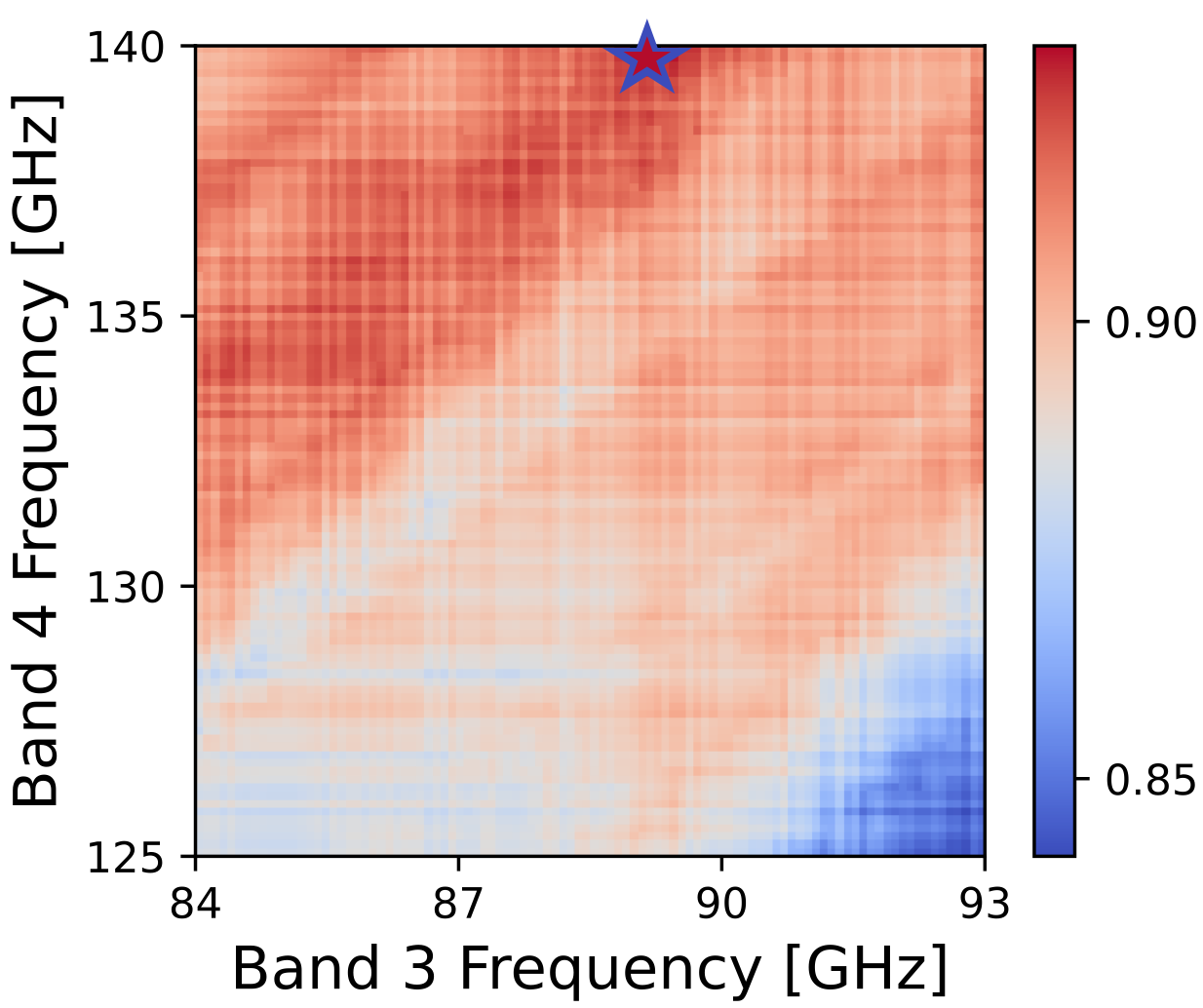}
    \caption{The distribution in redshift search capability of ALMA for different band 3 and 4 (3 and 2~mm, resp.) configurations for HerBS targets \citep{bakx18}. As a figure of merit, we take the robust redshift probability plus 50\% of the non-robust redshift probability.
    Here, we assume a 'stacked' configuration of three spectral windows in each of the respective bands in order to form two near-contiguous frequency ranges starting at the reported value. The optimum observing frequencies for such a combined band 3 and 4 observation with ALMA lie in the middle-frequency of the 3~mm window and at the upper-frequency end of the 2~mm window, as marked with the {\it red-and-blue star}. The frequency set-up in \citet{Urquhart2022} was derived using a similar method, and they found an efficiency in line with our findings ($
    \sim 72$\%, and up to $\sim92$\% with deeper observations). }
    \label{fig:band34_optimization}
\end{figure}
We test the ability to include 2~mm spectral windows to improve the robust redshift coverage of ALMA. The proposed configuration uses a total of six tunings to generate two near-contiguous frequency coverages at both 3 and 2~mm, which removes the overlapping region of the 3~mm scan. The total figure of merit is determined not solely by robust redshifts (either through single- or multiple-lines), but we choose to additionally add non-robust redshifts, weighted down by 50\%. 
Figure \ref{fig:band34_optimization} shows the figure of merit for each possible combination of 3 and 2~mm regions on a smoothed redshift distribution similar to the HerBS targets \citep{bakx18,Bakx2020Erratum}, assuming a low uncertainty in the $z_{\rm phot}$ of $\Delta z / (1+z) \approx 7$\%. Note that both the figure of merit, as well as the choice for the redshift distribution depend on the science case, and can be adjusted by the reader in the scripts provided online.
The figure of merit varies between 0.83 to 0.94 for different combinations of the 3 and 2~mm window, with the optimum lying at the middle frequencies in band 3 (3~mm) and at higher frequencies in band 4 (2~mm), with the precise configuration shown in Table~\ref{tab:efficiencies}. 
The frequencies of the ALMA 12m array observations in \cite{Urquhart2022} were derived using a similar method. They report redshift search efficiencies similar to the ones suggested by Figure~\ref{fig:band34_optimization}, with a robust redshift fraction of 72\%, and promising to go up to 93\% with deeper observations.
While we demonstrate the efficacy of this method on ALMA, any redshift search with tunable (sub)mm receivers can be optimised using this method through the publicly-available code.

\section{Conclusions}
This work documents new and efficient methods for (sub)mm redshift searches of high-$z$ SMGs. The tools discussed in this paper are publicly available at \url{https://github.com/tjlcbakx/redshift-search-graphs}. This paper reports on:
\begin{itemize}
\renewcommand\labelitemi{\textbf{ $\blacksquare{}$}}
    \item An efficient graphical method for identifying redshifts, removing the need to manually cross-compare redshift options. Aside from a description of the method, we provide a comprehensive overview of caveats in deriving redshifts using (sub)mm spectral lines in the high-$z$. The method provides highlights the explored redshift-space, enabling the removal of untenable redshift options, while also pointing out potential redshift degeneracies. 
    \item A method to calculate the capabilities of various instruments towards deriving robust spectroscopic redshifts. This is useful, not only to evaluate past missions, but also to guide future instruments.
    \item A method to optimise tunings towards robust spectroscopic redshifts (and other figures of merit) for ALMA and other tunable (sub)mm receivers. Here, we stress the need for both 3 and 2~mm observations to identify redshifts robustly.
\end{itemize}

\section*{Acknowledgements}
TB thanks the helpful contributions towards this publication from Dr. Ewoud A.I. Pool, who provided both a wider scope on the project and improved the graph descriptions. {\color{referee} We also thank the helpful comments of the referee, which increased the scientific scope of the paper and the associated code.}
TB acknowledges funding from NAOJ ALMA Scientific Research Grant Numbers 2018-09B and JSPS KAKENHI No. 17H06130. HD acknowledges financial support from the Spanish Ministry of Science, Innovation and Universities (MICIU) under the 2014 Ramón y Cajal program RYC-2014-15686, from the Agencia Estatal de Investigación del Ministerio de Ciencia e Innovación (AEI-MCINN) under grant (La evolución de los cíumulos de galaxias desde el amanecer hasta el mediodía cósmico) with reference (PID2019-105776GB-I00/DOI:10.13039/501100011033) and acknowledge support from the ACIISI, Consejería de Economía, Conocimiento y Empleo del Gobierno de Canarias and the European Regional Development Fund (ERDF) under grant with reference PROID2020010107. This paper makes use of the following ALMA data: ADS/JAO.ALMA 2016.2.00133.S, 2018.1.00804.S, 2019.1.01147 and 2019.2.00155.S.

\section*{Data Availability}
All code used to derive the graphs in this paper are available at \url{https://github.com/tjlcbakx/redshift-search-graphs} in order to assist in the robust classification of redshifts in ongoing, future and past programs.



\bibliographystyle{mnras}
\bibliography{example} 

\begin{thebibliography}{}
\makeatletter
\relax
\def\mn@urlcharsother{\let\do\@makeother \do\$\do\&\do\#\do\^\do\_\do\%\do\~}
\def\mn@doi{\begingroup\mn@urlcharsother \@ifnextchar [ {\mn@doi@}
  {\mn@doi@[]}}
\def\mn@doi@[#1]#2{\def\@tempa{#1}\ifx\@tempa\@empty \href
  {http://dx.doi.org/#2} {doi:#2}\else \href {http://dx.doi.org/#2} {#1}\fi
  \endgroup}
\def\mn@eprint#1#2{\mn@eprint@#1:#2::\@nil}
\def\mn@eprint@arXiv#1{\href {http://arxiv.org/abs/#1} {{\tt arXiv:#1}}}
\def\mn@eprint@dblp#1{\href {http://dblp.uni-trier.de/rec/bibtex/#1.xml}
  {dblp:#1}}
\def\mn@eprint@#1:#2:#3:#4\@nil{\def\@tempa {#1}\def\@tempb {#2}\def\@tempc
  {#3}\ifx \@tempc \@empty \let \@tempc \@tempb \let \@tempb \@tempa \fi \ifx
  \@tempb \@empty \def\@tempb {arXiv}\fi \@ifundefined
  {mn@eprint@\@tempb}{\@tempb:\@tempc}{\expandafter \expandafter \csname
  mn@eprint@\@tempb\endcsname \expandafter{\@tempc}}}

\bibitem[\protect\citeauthoryear{{Bakx} et~al.,}{{Bakx} et~al.}{2018}]{bakx18}
{Bakx} T. J.~L.~C.,  et~al., 2018, \mn@doi [\mnras] {10.1093/mnras/stx2267},
  \href {https://ui.adsabs.harvard.edu/abs/2018MNRAS.473.1751B} {473, 1751}

\bibitem[\protect\citeauthoryear{{Bakx}, {Eales}  \& {Amvrosiadis}}{{Bakx}
  et~al.}{2020a}]{bakx2020VIKING}
{Bakx} T. J.~L.~C.,  {Eales} S.,   {Amvrosiadis} A.,  2020a, \mn@doi [\mnras]
  {10.1093/mnras/staa506}, \href
  {https://ui.adsabs.harvard.edu/abs/2020MNRAS.493.4276B} {493, 4276}

\bibitem[\protect\citeauthoryear{{Bakx} et~al.,}{{Bakx}
  et~al.}{2020b}]{Bakx2020Erratum}
{Bakx} T. J.~L.~C.,  et~al., 2020b, \mn@doi [\mnras] {10.1093/mnras/staa658},
  \href {https://ui.adsabs.harvard.edu/abs/2020MNRAS.494...10B} {494, 10}

\bibitem[\protect\citeauthoryear{{Bakx} et~al.,}{{Bakx}
  et~al.}{2020c}]{Bakx2020IRAM}
{Bakx} T.~J.~L.~C.,  et~al., 2020c, \mn@doi [\mnras] {10.1093/mnras/staa1664},
  \href {https://ui.adsabs.harvard.edu/abs/2020MNRAS.496.2372B} {496, 2372}

\bibitem[\protect\citeauthoryear{{Battersby} et~al.,}{{Battersby}
  et~al.}{2018}]{Battersby2018}
{Battersby} C.,  et~al., 2018, \mn@doi [Nature Astronomy]
  {10.1038/s41550-018-0540-y}, \href
  {https://ui.adsabs.harvard.edu/abs/2018NatAs...2..596B} {2, 596}

\bibitem[\protect\citeauthoryear{{Baugh}, {Lacey}, {Frenk}, {Granato}, {Silva},
  {Bressan}, {Benson}  \& {Cole}}{{Baugh} et~al.}{2005}]{Baugh2005}
{Baugh} C.~M.,  {Lacey} C.~G.,  {Frenk} C.~S.,  {Granato} G.~L.,  {Silva} L.,
  {Bressan} A.,  {Benson} A.~J.,   {Cole} S.,  2005, \mn@doi [\mnras]
  {10.1111/j.1365-2966.2004.08553.x}, \href
  {https://ui.adsabs.harvard.edu/abs/2005MNRAS.356.1191B} {356, 1191}

\bibitem[\protect\citeauthoryear{{Blain}, {Smail}, {Ivison}, {Kneib}  \&
  {Frayer}}{{Blain} et~al.}{2002}]{blain2002}
{Blain} A.~W.,  {Smail} I.,  {Ivison} R.~J.,  {Kneib} J.~P.,   {Frayer} D.~T.,
  2002, \mn@doi [\physrep] {10.1016/S0370-1573(02)00134-5}, \href
  {https://ui.adsabs.harvard.edu/abs/2002PhR...369..111B} {369, 111}

\bibitem[\protect\citeauthoryear{{Blain}, {Barnard}  \& {Chapman}}{{Blain}
  et~al.}{2003}]{Blain2003}
{Blain} A.~W.,  {Barnard} V.~E.,   {Chapman} S.~C.,  2003, \mn@doi [\mnras]
  {10.1046/j.1365-8711.2003.06086.x}, \href
  {https://ui.adsabs.harvard.edu/abs/2003MNRAS.338..733B} {338, 733}

\bibitem[\protect\citeauthoryear{{Bradford} et~al.,}{{Bradford}
  et~al.}{2021}]{Bradford2021}
{Bradford} C.~M.,  et~al., 2021, \mn@doi [Journal of Astronomical Telescopes,
  Instruments, and Systems] {10.1117/1.JATIS.7.1.011017}, \href
  {https://ui.adsabs.harvard.edu/abs/2021JATIS...7a1017B} {7, 011017}

\bibitem[\protect\citeauthoryear{{Carilli} \& {Walter}}{{Carilli} \&
  {Walter}}{2013}]{carilli2013}
{Carilli} C.~L.,  {Walter} F.,  2013, \mn@doi [\araa]
  {10.1146/annurev-astro-082812-140953}, \href
  {https://ui.adsabs.harvard.edu/abs/2013ARA&A..51..105C} {51, 105}

\bibitem[\protect\citeauthoryear{{Casey}}{{Casey}}{2020}]{Casey2020}
{Casey} C.~M.,  2020, \mn@doi [\apj] {10.3847/1538-4357/aba528}, \href
  {https://ui.adsabs.harvard.edu/abs/2020ApJ...900...68C} {900, 68}

\bibitem[\protect\citeauthoryear{{Casey} et~al.,}{{Casey}
  et~al.}{2012}]{casey2012}
{Casey} C.~M.,  et~al., 2012, \mn@doi [\apj] {10.1088/0004-637X/761/2/140},
  \href {https://ui.adsabs.harvard.edu/abs/2012ApJ...761..140C} {761, 140}

\bibitem[\protect\citeauthoryear{{Casey}, {Narayanan}  \& {Cooray}}{{Casey}
  et~al.}{2014}]{casey2014}
{Casey} C.~M.,  {Narayanan} D.,   {Cooray} A.,  2014, \mn@doi [\physrep]
  {10.1016/j.physrep.2014.02.009}, \href
  {https://ui.adsabs.harvard.edu/abs/2014PhR...541...45C} {541, 45}

\bibitem[\protect\citeauthoryear{{Casey} et~al.,}{{Casey}
  et~al.}{2018}]{Casey2018}
{Casey} C.~M.,  et~al., 2018, \mn@doi [\apj] {10.3847/1538-4357/aac82d}, \href
  {https://ui.adsabs.harvard.edu/abs/2018ApJ...862...77C} {862, 77}

\bibitem[\protect\citeauthoryear{{Chapman}, {Blain}, {Smail}  \&
  {Ivison}}{{Chapman} et~al.}{2005}]{chapman2005}
{Chapman} S.~C.,  {Blain} A.~W.,  {Smail} I.,   {Ivison} R.~J.,  2005, \mn@doi
  [\apj] {10.1086/428082}, \href
  {https://ui.adsabs.harvard.edu/abs/2005ApJ...622..772C} {622, 772}

\bibitem[\protect\citeauthoryear{{Chen} et~al.,}{{Chen}
  et~al.}{2021}]{Chen2022}
{Chen} C.-C.,  et~al., 2021, arXiv e-prints, \href
  {https://ui.adsabs.harvard.edu/abs/2021arXiv211207430C} {p. arXiv:2112.07430}

\bibitem[\protect\citeauthoryear{{Day}, {LeDuc}, {Mazin}, {Vayonakis}  \&
  {Zmuidzinas}}{{Day} et~al.}{2003}]{Day2003}
{Day} P.~K.,  {LeDuc} H.~G.,  {Mazin} B.~A.,  {Vayonakis} A.,   {Zmuidzinas}
  J.,  2003, \mn@doi [\nat] {10.1038/nature02037}, \href
  {https://ui.adsabs.harvard.edu/abs/2003Natur.425..817D} {425, 817}

\bibitem[\protect\citeauthoryear{{Donevski} et~al.,}{{Donevski}
  et~al.}{2018}]{Donevski2018}
{Donevski} D.,  et~al., 2018, \mn@doi [\aap] {10.1051/0004-6361/201731888},
  \href {https://ui.adsabs.harvard.edu/abs/2018A&A...614A..33D} {614, A33}

\bibitem[\protect\citeauthoryear{{Duivenvoorden} et~al.,}{{Duivenvoorden}
  et~al.}{2018}]{Duivenvoorden2018}
{Duivenvoorden} S.,  et~al., 2018, \mn@doi [\mnras] {10.1093/mnras/sty691},
  \href {https://ui.adsabs.harvard.edu/abs/2018MNRAS.477.1099D} {477, 1099}

\bibitem[\protect\citeauthoryear{{Endo} et~al.,}{{Endo}
  et~al.}{2019}]{Endo2019}
{Endo} A.,  et~al., 2019, \mn@doi [Nature Astronomy]
  {10.1038/s41550-019-0850-8}, \href
  {https://ui.adsabs.harvard.edu/abs/2019NatAs...3..989E} {3, 989}

\bibitem[\protect\citeauthoryear{{Erickson}, {Narayanan}, {Goeller}  \&
  {Grosslein}}{{Erickson} et~al.}{2007}]{Erickson2007}
{Erickson} N.,  {Narayanan} G.,  {Goeller} R.,   {Grosslein} R.,  2007, in
  {Baker} A.~J.,  {Glenn} J.,  {Harris} A.~I.,  {Mangum} J.~G.,   {Yun} M.~S.,
  eds,  Astronomical Society of the Pacific Conference Series Vol. 375, From
  Z-Machines to ALMA: (Sub)Millimeter Spectroscopy of Galaxies. p.~71

\bibitem[\protect\citeauthoryear{{Fardal}, {Katz}, {Weinberg}, {Dav{\'e}}  \&
  {Hernquist}}{{Fardal} et~al.}{2001}]{Fardal2001}
{Fardal} M.~A.,  {Katz} N.,  {Weinberg} D.~H.,  {Dav{\'e}} R.,   {Hernquist}
  L.,  2001, arXiv e-prints, \href
  {https://ui.adsabs.harvard.edu/abs/2001astro.ph..7290F} {pp
  astro--ph/0107290}

\bibitem[\protect\citeauthoryear{{Fixsen}, {Cheng}, {Gales}, {Mather}, {Shafer}
   \& {Wright}}{{Fixsen} et~al.}{1996}]{Fixsen1996}
{Fixsen} D.~J.,  {Cheng} E.~S.,  {Gales} J.~M.,  {Mather} J.~C.,  {Shafer}
  R.~A.,   {Wright} E.~L.,  1996, \mn@doi [\apj] {10.1086/178173}, \href
  {https://ui.adsabs.harvard.edu/abs/1996ApJ...473..576F} {473, 576}

\bibitem[\protect\citeauthoryear{{Frayer} et~al.,}{{Frayer}
  et~al.}{2011}]{Frayer2011}
{Frayer} D.~T.,  et~al., 2011, \mn@doi [\apjl] {10.1088/2041-8205/726/2/L22},
  \href {https://ui.adsabs.harvard.edu/abs/2011ApJ...726L..22F} {726, L22}

\bibitem[\protect\citeauthoryear{{Gonz{\'a}lez}, {Labb{\'e}}, {Bouwens},
  {Illingworth}, {Franx}  \& {Kriek}}{{Gonz{\'a}lez}
  et~al.}{2011}]{Gonzalez2011}
{Gonz{\'a}lez} V.,  {Labb{\'e}} I.,  {Bouwens} R.~J.,  {Illingworth} G.,
  {Franx} M.,   {Kriek} M.,  2011, \mn@doi [\apjl]
  {10.1088/2041-8205/735/2/L34}, \href
  {https://ui.adsabs.harvard.edu/abs/2011ApJ...735L..34G} {735, L34}

\bibitem[\protect\citeauthoryear{{Harrington} et~al.,}{{Harrington}
  et~al.}{2021}]{Harrington2021}
{Harrington} K.~C.,  et~al., 2021, \mn@doi [\apj] {10.3847/1538-4357/abcc01},
  \href {https://ui.adsabs.harvard.edu/abs/2021ApJ...908...95H} {908, 95}

\bibitem[\protect\citeauthoryear{{Harris} et~al.,}{{Harris}
  et~al.}{2007}]{Harris2007}
{Harris} A.~I.,  et~al., 2007, in {Baker} A.~J.,  {Glenn} J.,  {Harris} A.~I.,
  {Mangum} J.~G.,   {Yun} M.~S.,  eds,  Astronomical Society of the Pacific
  Conference Series Vol. 375, From Z-Machines to ALMA: (Sub)Millimeter
  Spectroscopy of Galaxies. p.~82

\bibitem[\protect\citeauthoryear{{Harris}, {Baker}, {Zonak}, {Sharon},
  {Genzel}, {Rauch}, {Watts}  \& {Creager}}{{Harris} et~al.}{2010}]{harris2010}
{Harris} A.~I.,  {Baker} A.~J.,  {Zonak} S.~G.,  {Sharon} C.~E.,  {Genzel} R.,
  {Rauch} K.,  {Watts} G.,   {Creager} R.,  2010, \mn@doi [\apj]
  {10.1088/0004-637X/723/2/1139}, \href
  {https://ui.adsabs.harvard.edu/abs/2010ApJ...723.1139H} {723, 1139}

\bibitem[\protect\citeauthoryear{{Harris} et~al.,}{{Harris}
  et~al.}{2012}]{harris2012}
{Harris} A.~I.,  et~al., 2012, \mn@doi [\apj] {10.1088/0004-637X/752/2/152},
  \href {https://ui.adsabs.harvard.edu/abs/2012ApJ...752..152H} {752, 152}

\bibitem[\protect\citeauthoryear{{Hughes} et~al.,}{{Hughes}
  et~al.}{1998}]{Hughes1998}
{Hughes} D.~H.,  et~al., 1998, \mn@doi [\nat] {10.1038/28328}, \href
  {https://ui.adsabs.harvard.edu/abs/1998Natur.394..241H} {394, 241}

\bibitem[\protect\citeauthoryear{{Ivison} et~al.,}{{Ivison}
  et~al.}{2016}]{ivison16}
{Ivison} R.~J.,  et~al., 2016, \mn@doi [\apj] {10.3847/0004-637X/832/1/78},
  \href {https://ui.adsabs.harvard.edu/abs/2016ApJ...832...78I} {832, 78}

\bibitem[\protect\citeauthoryear{{Lacey}, {Baugh}, {Frenk}, {Silva}, {Granato}
  \& {Bressan}}{{Lacey} et~al.}{2008}]{Lacey2008}
{Lacey} C.~G.,  {Baugh} C.~M.,  {Frenk} C.~S.,  {Silva} L.,  {Granato} G.~L.,
  {Bressan} A.,  2008, \mn@doi [\mnras] {10.1111/j.1365-2966.2008.12949.x},
  \href {https://ui.adsabs.harvard.edu/abs/2008MNRAS.385.1155L} {385, 1155}

\bibitem[\protect\citeauthoryear{{Lupu} et~al.,}{{Lupu}
  et~al.}{2012}]{lupu2012}
{Lupu} R.~E.,  et~al., 2012, \mn@doi [\apj] {10.1088/0004-637X/757/2/135},
  \href {https://ui.adsabs.harvard.edu/abs/2012ApJ...757..135L} {757, 135}

\bibitem[\protect\citeauthoryear{{Marrone} et~al.,}{{Marrone}
  et~al.}{2018}]{Marrone2018}
{Marrone} D.~P.,  et~al., 2018, \mn@doi [\nat] {10.1038/nature24629}, \href
  {https://ui.adsabs.harvard.edu/abs/2018Natur.553...51M} {553, 51}

\bibitem[\protect\citeauthoryear{{Narayanan} et~al.,}{{Narayanan}
  et~al.}{2015}]{narayanan2015}
{Narayanan} D.,  et~al., 2015, \mn@doi [\nat] {10.1038/nature15383}, \href
  {https://ui.adsabs.harvard.edu/abs/2015Natur.525..496N} {525, 496}

\bibitem[\protect\citeauthoryear{{Naylor} et~al.,}{{Naylor}
  et~al.}{2003}]{Naylor2003}
{Naylor} B.~J.,  et~al., 2003, in {Phillips} T.~G.,  {Zmuidzinas} J.,  eds,
  Society of Photo-Optical Instrumentation Engineers (SPIE) Conference Series
  Vol. 4855, Millimeter and Submillimeter Detectors for Astronomy. pp 239--248,
  \mn@doi{10.1117/12.459419}

\bibitem[\protect\citeauthoryear{{Neri} et~al.,}{{Neri}
  et~al.}{2020}]{neri2019}
{Neri} R.,  et~al., 2020, \mn@doi [\aap] {10.1051/0004-6361/201936988}, \href
  {https://ui.adsabs.harvard.edu/abs/2020A&A...635A...7N} {635, A7}

\bibitem[\protect\citeauthoryear{{Pearson} et~al.,}{{Pearson}
  et~al.}{2013}]{pearson13}
{Pearson} E.~A.,  et~al., 2013, \mn@doi [\mnras] {10.1093/mnras/stt1369}, \href
  {https://ui.adsabs.harvard.edu/abs/2013MNRAS.435.2753P} {435, 2753}

\bibitem[\protect\citeauthoryear{{Redford} et~al.,}{{Redford}
  et~al.}{2021}]{Redford2021}
{Redford} J.,  et~al., 2021, in American Astronomical Society Meeting
  Abstracts. p. 317.01

\bibitem[\protect\citeauthoryear{{Reuter} et~al.,}{{Reuter}
  et~al.}{2020}]{reuter20}
{Reuter} C.,  et~al., 2020, \mn@doi [\apj] {10.3847/1538-4357/abb599}, \href
  {https://ui.adsabs.harvard.edu/abs/2020ApJ...902...78R} {902, 78}

\bibitem[\protect\citeauthoryear{{Rowan-Robinson} et~al.,}{{Rowan-Robinson}
  et~al.}{2016}]{RowanRobinson2016}
{Rowan-Robinson} M.,  et~al., 2016, \mn@doi [\mnras] {10.1093/mnras/stw1169},
  \href {https://ui.adsabs.harvard.edu/abs/2016MNRAS.461.1100R} {461, 1100}

\bibitem[\protect\citeauthoryear{{Rybak} et~al.,}{{Rybak}
  et~al.}{2021}]{Rybak2021}
{Rybak} M.,  et~al., 2021, arXiv e-prints, \href
  {https://ui.adsabs.harvard.edu/abs/2021arXiv211105261R} {p. arXiv:2111.05261}

\bibitem[\protect\citeauthoryear{{Smail}, {Ivison}  \& {Blain}}{{Smail}
  et~al.}{1997}]{Smail1997}
{Smail} I.,  {Ivison} R.~J.,   {Blain} A.~W.,  1997, \mn@doi [\apjl]
  {10.1086/311017}, \href
  {https://ui.adsabs.harvard.edu/abs/1997ApJ...490L...5S} {490, L5}

\bibitem[\protect\citeauthoryear{{Strandet} et~al.,}{{Strandet}
  et~al.}{2016}]{strandet2016}
{Strandet} M.~L.,  et~al., 2016, \mn@doi [\apj] {10.3847/0004-637X/822/2/80},
  \href {https://ui.adsabs.harvard.edu/abs/2016ApJ...822...80S} {822, 80}

\bibitem[\protect\citeauthoryear{{Taniguchi} et~al.,}{{Taniguchi}
  et~al.}{2021}]{Taniguchi2021}
{Taniguchi} A.,  et~al., 2021, arXiv e-prints, \href
  {https://ui.adsabs.harvard.edu/abs/2021arXiv211014656T} {p. arXiv:2110.14656}

\bibitem[\protect\citeauthoryear{{Urquhart} et~al.,}{{Urquhart}
  et~al.}{2022}]{Urquhart2022}
{Urquhart} S.~A.,  et~al., 2022, arXiv e-prints, \href
  {https://ui.adsabs.harvard.edu/abs/2022arXiv220107815U} {p. arXiv:2201.07815}

\bibitem[\protect\citeauthoryear{{Valentino} et~al.,}{{Valentino}
  et~al.}{2018}]{Valentino2018}
{Valentino} F.,  et~al., 2018, \mn@doi [\apj] {10.3847/1538-4357/aaeb88}, \href
  {https://ui.adsabs.harvard.edu/abs/2018ApJ...869...27V} {869, 27}

\bibitem[\protect\citeauthoryear{{Vieira} et~al.,}{{Vieira}
  et~al.}{2013}]{Vieira2013}
{Vieira} J.~D.,  et~al., 2013, \mn@doi [\nat] {10.1038/nature12001}, \href
  {https://ui.adsabs.harvard.edu/abs/2013Natur.495..344V} {495, 344}

\bibitem[\protect\citeauthoryear{{Wei{\ss}} et~al.,}{{Wei{\ss}}
  et~al.}{2009}]{weiss2009}
{Wei{\ss}} A.,  et~al., 2009, \mn@doi [\apj] {10.1088/0004-637X/707/2/1201},
  \href {https://ui.adsabs.harvard.edu/abs/2009ApJ...707.1201W} {707, 1201}

\bibitem[\protect\citeauthoryear{{Wei{\ss}} et~al.,}{{Wei{\ss}}
  et~al.}{2013}]{weiss2013}
{Wei{\ss}} A.,  et~al., 2013, \mn@doi [\apj] {10.1088/0004-637X/767/1/88},
  \href {https://ui.adsabs.harvard.edu/abs/2013ApJ...767...88W} {767, 88}

\bibitem[\protect\citeauthoryear{Xu et~al.,}{Xu et~al.}{2014}]{Xu2014}
Xu C.,  et~al., 2014, \mn@doi [The Astrophysical Journal]
  {10.1088/0004-637X/787/1/48}, 787

\bibitem[\protect\citeauthoryear{Yanai \& Lercher}{Yanai \&
  Lercher}{2020}]{Yanai2020}
Yanai I.,  Lercher M.,  2020, \mn@doi [Genome Biology]
  {10.1186/s13059-020-02133-w}, 21, 231

\bibitem[\protect\citeauthoryear{{Zavala} et~al.,}{{Zavala}
  et~al.}{2015}]{Zavala2015}
{Zavala} J.~A.,  et~al., 2015, \mn@doi [\mnras] {10.1093/mnras/stv1351}, \href
  {https://ui.adsabs.harvard.edu/abs/2015MNRAS.452.1140Z} {452, 1140}

\bibitem[\protect\citeauthoryear{{Zavala} et~al.,}{{Zavala}
  et~al.}{2018}]{Zavala2018}
{Zavala} J.~A.,  et~al., 2018, \mn@doi [Nature Astronomy]
  {10.1038/s41550-017-0297-8}, \href
  {https://ui.adsabs.harvard.edu/abs/2018NatAs...2...56Z} {2, 56}

\bibitem[\protect\citeauthoryear{{da Cunha} et~al.,}{{da Cunha}
  et~al.}{2015}]{cunha2015ALESS}
{da Cunha} E.,  et~al., 2015, \mn@doi [\apj] {10.1088/0004-637X/806/1/110},
  \href {https://ui.adsabs.harvard.edu/abs/2015ApJ...806..110D} {806, 110}

\makeatother
\end{thebibliography}




\bsp	
\label{lastpage}
\end{document}